\documentclass[journal]{IEEEtran}
\IEEEoverridecommandlockouts
\usepackage{cite}
\usepackage{amsmath,amssymb,amsfonts}
\usepackage{algorithmic}
\usepackage{graphicx}
\usepackage{textcomp}
\usepackage{xcolor}
\usepackage{amsthm}
\usepackage{bm}
\usepackage{hyperref}
\usepackage{balance}

\def\BibTeX{{\rm B\kern-.05em{\sc i\kern-.025em b}\kern-.08em
    T\kern-.1667em\lower.7ex\hbox{E}\kern-.125emX}}

\newcommand{\X}{\mathcal{X}}
\newcommand{\Y}{\mathcal{Y}}
\newcommand{\N}{\mathcal{N}}
\newcommand{\U}{\mathcal{U}}

\newcommand{\G}{\mathcal{G}}
\newcommand{\E}{\mathbb{E}}
\newcommand{\R}{\mathbb{R}}
\newcommand{\Z}{\mathbb{Z}}
\newcommand{\F}{\mathcal{F}}
\newcommand{\LL}{\mathcal{L}}
\newcommand{\Prob}{\mathbb{P}}

\newcommand{\Brace}[1]{\left\{#1\right\}}
\newcommand{\Paren}[1]{\left(#1\right)}
\newcommand{\Br}[1]{\left[#1\right]}

\DeclareMathOperator{\var}{var}
\DeclareMathOperator{\st}{s. t.}
\DeclareMathOperator*{\argmin}{arg\,min}

\newtheorem{theorem}{Theorem}
\newtheorem{remark}{Remark}
\newtheorem{lemma}{Lemma}

\begin{document}

\title{Variable-Length Stop-Feedback Codes With Finite Optimal Decoding Times for BI-AWGN Channels
\thanks{H. Yang and R. D. Wesel are with the Department of Electrical and Computer Engineering, University of California, Los Angeles (e-mail: \{hengjie.yang,  wesel\}@ucla.edu). R. C. Yavas and V. Kostina are with the Department of Electrical Engineering, California Institute of Technology (e-mail: \{ryavas,  vkostina\}@caltech.edu). This research is supported by National Science Foundation (NSF) grant CCF-1955660.}
}
\author{\IEEEauthorblockN{
Hengjie Yang, Recep Can Yavas, Victoria Kostina, and Richard D. Wesel}
}


\maketitle 

\begin{abstract}
In this paper, we are interested in the performance of a variable-length stop-feedback (VLSF) code with $m$ optimal decoding times for the binary-input additive white Gaussian noise channel. We first develop tight approximations on the tail probability of length-$n$ cumulative information density. Building on the work of Yavas \emph{et al.}, for a given information density threshold, we formulate the integer program of minimizing the upper bound on average blocklength over all decoding times subject to the average error probability, minimum gap and integer constraints. Eventually, minimization of locally minimum upper bounds over all thresholds will yield the globally minimum upper bound and this is called the two-step minimization. For the integer program, we present a greedy algorithm that yields possibly suboptimal integer decoding times. By allowing a positive real-valued decoding time, we develop the gap-constrained sequential differential optimization (SDO) procedure that sequentially produces the optimal,  real-valued decoding times. We identify the error regime in which Polyanskiy's scheme of stopping at zero does not improve the achievability bound. In this error regime, the two-step minimization with the gap-constrained SDO shows that a finite $m$ suffices to attain Polyanskiy's bound for VLSF codes with $m = \infty$.
\end{abstract}


\section{Introduction}
Feedback has been shown to be useful both in the variable-length and fixed-length regimes, even though it does not improve the capacity of a memoryless, point-to-point channel \cite{Shannon1956}. In the variable-length regime, feedback has been shown to simplify the construction of coding schemes \cite{Horstein1963,Schalkwijk1966,Shayevitz2011}, to significantly improve the optimal error exponent \cite{Burnashev1976}, and to achieve universality \cite{Luby2002,Draper2004,Yavas_TIT2021}. In the fixed-length regime, feedback is shown to improve the second-order coding rate for the compound-dispersion discrete memoryless channels \cite{Wagner2020}.

In \cite{Polyanskiy2011}, Polyanskiy \emph{et al.} introduced variable-length feedback (VLF) codes, variable-length feedback with termination (VLFT) codes, and a special VLF code called a \emph{variable-length stop-feedback (VLSF)} code. The infinite-length VLSF codewords are fixed before the start of transmission and feedback only affects the portion of a codeword being transmitted rather than the value of that codeword. During transmission, a feedback symbol ``0'' indicates that the decoder is not ready to decode and the transmission should continue, whereas a ``1'' signifies that the decoder is ready to decode and the transmitter must stop. Using VLSF codes, Polyanskiy \emph{et al.} demonstrated that $\frac{C}{1-\epsilon}$ is achievable by stopping the code at $\tau=0$ with a small probability, where $C$ denotes channel capacity, and $\epsilon$ denotes the target error probability \cite{Polyanskiy2011}.

The VLSF code defined in \cite{Polyanskiy2011} can be thought of as a VLF code with infinitely many decoding times, i.e., the number of decoding times $m = \infty$. However, in practical systems, the feedback opportunities are limited, i.e., $m < \infty$, and the decoder is only allowed to decode at time instants $n_1, n_2, \dots, n_m$. In \cite{Kim2015}, Kim \emph{et al.} investigated VLSF codes with $m$ periodic decoding times and derived a lower bound on throughput.  In order to minimize the average blocklength, Vakilinia \emph{et al.} \cite{Vakilinia2016} developed the \emph{sequential differential optimization} (SDO) algorithm that produces decoding time $n_{k+1}$ based on the knowledge of $n_{k}$, $n_{k-1}$, and their successful decoding probabilities approximated by a differentiable function. The  SDO in \cite{Vakilinia2016} uses the Gaussian tail probability to approximate the probability of successful decoding. Later, variations of SDO were developed to improve the Gaussian model accuracy \cite{Wang2017, Wesel2018}. The SDO algorithm is used to optimize systems that employ incremental redundancy and hybrid automatic repeat request (ARQ) \cite{Wong2017}, and to code for the binary erasure channel \cite{Heidarzadeh2018, Heidarzadeh2019}. However, in this paper, we show that the Gaussian model is still imprecise for small values of $n$. Additionally, the existing SDO procedure fails to consider the inherent gap constraint that two decoding times must be separated by at least one.

In \cite{Yavas2021}, Yavas \emph{et al.} developed an achievability bound for VLSF codes with $m$ decoding times for the additive white Gaussian noise channel with capacity $C$, dispersion $V$, and maximal power constraint $P$. The asymptotic expansion of the maximum message size $M$ is given by $\ln M \approx \frac{lC}{1-\epsilon} - \sqrt{l\ln_{(m-1)}(l)\frac{V}{1-\epsilon}}$ where $\ln_{(k)}(\cdot)$ denotes the $k$-fold nested logarithm, $l$ and $\epsilon$ are the upper bounds on average blocklength and error probability of the VLSF code, respectively. They showed that a slight increase in $m$ can dramatically improve the achievable rate of VLSF codes. Unfortunately, due to the nested logarithm term, Yavas \emph{et al.} were only able to show achievability bounds for $m\le 4$ for average blocklength less than $2000$. They also demonstrated that within their code construction, the decoding times chosen by the SDO will yield the same second- and third-order coding rates as attained by their construction of decoding times.

In this paper, we are interested in the performance of a VLSF code with $m$ optimal decoding times for the binary-input additive white Gaussian noise (BI-AWGN) channel. We first develop tight approximations on the tail probability of length-$n$ cumulative information density. Building on the result of Yavas \emph{et al.} \cite{Yavas2021}, for a fixed information density threshold $\gamma$, we formulate an integer program of minimizing the upper bound on average blocklength over all decoding times $n_1, n_2, \dots, n_m$ subject to average error probability, minimum gap and integer constraints. Finally, minimization of locally minimum upper bounds over information density threshold $\gamma$ yields the globally minimum upper bound, and this method is called the \emph{two-step minimization}. For the integer program, we present a greedy algorithm that yields possibly suboptimal integer decoding times. By allowing positive real-valued decoding times, we develop the \emph{gap-constrained SDO} algorithm that captures the minimum gap constraint for the relaxed program. In \cite{Polyanskiy2011}, Polyanskiy \emph{et al.} demonstrated that the rate $\frac{C}{1-\epsilon}$ is achievable by allowing the VLSF code to stop at zero with a small probability. In this paper, we identify the error regime where Polyanskiy's scheme of stopping at zero does not improve the achievability bound. In this error regime, the two-step minimization with the gap-constrained SDO shows that a finite $m$ suffices to attain Polyanskiy's bound for VLSF codes with $m = \infty$.

This paper is organized as follows. Section \ref{sec: preliminaries} introduces the notation, the BI-AWGN channel model, and the VLSF code with $m$ decoding times. Section \ref{sec: approximation} develops tight approximations on the tail probability of length-$n$ cumulative information density. Section \ref{sec: SDO for VLSF codes} introduces the integer program, the two-step minimization, and a greedy algorithm, develops the gap-constrained SDO procedure for the relaxed program, identifies the error regime where stopping at zero does not help, and shows numerical comparisons. Section \ref{sec: conclusion} concludes the paper.

\section{Preliminaries}\label{sec: preliminaries}

\subsection{Notation}
For $k\in\Z_+$, $[k]\triangleq \{1,2,\dots, k\}$. We use $x_i^j$ to denote a sequence $(x_i, x_{i+1},\dots, x_j)$, $1\le i\le j$. When the context is clear, $x_1^n$ is abbreviated as $x^n$. All logarithms are taken to the base $2$. We use $\phi(x), \Phi(x), Q(\cdot)$ to respectively denote the probability density function (PDF), cumulative distribution function (CDF), and the tail probability of a standard normal $\N(0, 1)$.

\subsection{Channel Model and VLSF Codes with $m$ Decoding Times}

Let $X^n$ be a sequence of independent and identically distributed (i.i.d.) random variables, with each $X_i$ uniformly distributed over $\{-1, 1\}$. The output $Y^n$ of a memoryless, point-to-point BI-AWGN channel in response to $X^n$ is given by
\begin{align}
    Y^n = \sqrt{P}X^n + Z^n, \label{eq: BI-AWGN}
\end{align}
where $\sqrt{P}$ denotes the amplitude of binary-phase shift keying (BPSK), and $Z_1, Z_2,\dots, Z_n$ are i.i.d. standard normal random variables. The SNR of the BI-AWGN channel is given by $P$. 

For a BI-AWGN channel with a uniformly distributed input symbol, the information density $\iota(x; y)\triangleq \log\frac{P(y|x)}{P(y)}$ is given by
\begin{align}
    \iota(x; y) &= 1 - \log\Paren{1 + \exp\big(-2xy\sqrt{P}\big) }.
\end{align}
Since the channel is memoryless, the cumulative information density for $x^n$ and $y^n$ is given by
\begin{align}
    \iota(x^n; y^n) \triangleq \log\frac{P(y^n|x^n)}{P(y^n)} =  \sum_{i=1}^n\iota(x_i; y_i).
\end{align}
For a BI-AWGN channel, the channel capacity $C = \E[\iota(X;Y)]$ and dispersion $V = \var(\iota(X;Y))$.

Next, we follow \cite{Yavas2021} in describing a VLSF code with $m$ decoding times for the BI-AWGN channel. Due to BPSK, we omit the power constraint from the definition.

An $(l, n_1^m, M, \epsilon)$ VLSF code, where $l$ is a positive real, $n_1^m$ and $M$ are non-negative integers satisfying $n_1<n_2<\cdots<n_m$, $\epsilon\in(0, 1) $, is defined by
\begin{itemize}
  \item[1)] A finite alphabet $\U$ and a probability distribution $P_U$ on $\U$ defining the common randomness random variable $U$ that is revealed to both the transmitter and the receiver before the start of transmission.
  \item[2)] A sequence of encoders $f_n: \U\times[M]\to \X $, $n=1,2,\dots, n_m$, defining channel inputs
    \begin{align}
      X_n = f_n(U, W),
    \end{align}
    where $W\in[M]$ is the equiprobable message.
  \item[3)] A non-negative integer-valued random stopping time $\tau\in\{n_1, n_2, \dots, n_m\}$ of the filtration generated by $\G_m = \sigma\{U, Y^{n_m}\}$ that satisfies an average decoding time constraint
    \begin{align}
        \E[\tau] \le l. \label{eq: blocklength constraint}
    \end{align}
  \item[4)] $m$ decoding functions $g_{n_i}:\U\times\Y^{n_i}\to [M] $, providing the best estimate of $W$ at time $n_i$, $i=1,2,\dots,m$. The final decision $\hat{W}$ is computed at time instant $\tau$, i.e., $\hat{W} = g_{\tau}(U, Y^{\tau})$ and must satisfy
    \begin{align}
      P_e \triangleq \Prob[\hat{W} \ne W] \le \epsilon. \label{eq: error prob constraint}
    \end{align}
\end{itemize}
The rate of a VLSF code is given by $R \triangleq \log M/\E[\tau]$. In the above definition, the cardinality $\U$ specifies the number of deterministic codes under consideration to construct the random code. In \cite[Appendix D]{Yavas_TIT2021}, Yavas \emph{et al.} showed that $|\U|\le2$ suffices.

\section{Tight Approximations on $\Prob[\iota(X^{n}; Y^{n}) \ge \gamma]$}\label{sec: approximation}
In the analysis of $(l, n_1^m, M, \epsilon)$ VLSF codes, a key step is to develop a differentiable function $F_{\gamma}(n)$ to approximate or to bound the tail probability $\Prob[\iota(X^{n}; Y^{n}) \ge \gamma]$ with a fixed $\gamma$. In \cite{Vakilinia2016,Wang2017,Wong2017,Wesel2018}, $\Prob[\iota(X^{n}; Y^{n}) \ge \gamma]$ is approximated as a Gaussian tail probability, e.g., $Q\Paren{\frac{\gamma - nC}{\sqrt{nV}}}$ used in \cite{Wang2017}. However, we will show that for short blocklength $n$, the Gaussian model is imprecise and a better approximation is desired.

In probability theory, the Edgeworth expansion \cite{Edgeworth1905} has been known as a powerful tool to approximate the distribution of the sum of $n$ i.i.d. random variables. In this paper, we apply the order-$s$ Edgeworth expansion to approximate $\Prob[\iota(X^{n}; Y^{n}) \ge \gamma]$ for moderate and large values of $n$. We refer the reader to \cite[Chapter 2]{Peter1992} for a detailed introduction. 

\begin{theorem}[Equation (2.18), \cite{Peter1992}]\label{theorem: 3}
    Let $W_1, W_2, \dots, W_n$ be a sequence of i.i.d. random variables with zero mean and a finite variance $\sigma^2$. Define $G_n(x) \triangleq\Prob[\sum_{i=1}^nW_i \le x\sigma\sqrt{n}]$. Let $\chi_W(t)\triangleq \E[e^{itW}]$ be the characteristic function of $W$. If $\E[|W|^{s+2}] < \infty$ for some $s\in\Z^+$ and $\limsup_{|t|\to\infty}|\chi_W(t)| < 1$ (known as Cram\'er's condition), then,
    \begin{align}
    G_n(x) = \Phi(x) + \phi(x)\sum_{j=1}^{s}n^{-\frac{j}{2}}p_j(x) + o(n^{-s/2}),
    \end{align}
    where 
    \begin{align}
        &p_j(x) {=} -\sum_{\Brace{k_m}}He_{j+2r-1}(x)\prod_{m=1}^j\frac{1}{k_m!}\Paren{\frac{\kappa_{m+2}}{(m+2)!} }^{k_m}, \label{eq: 15}  \\
        &He_j(x) = j!\sum_{k=0}^{\lfloor j/2 \rfloor }\frac{(-1)^kx^{j-2k}}{k!(j-2k)!2^k}, \label{eq: 16}\\
        &\kappa_m = m!\sum_{\Brace{k_l}}(-1)^{r-1}(r-1)!\prod_{l=1}^m\frac{1}{k_l!}\Paren{\frac{\E[W^l]}{\sigma^l l!} }^{k_l}, \label{eq: 17}
    \end{align}
    where in \eqref{eq: 15}, the set $\Brace{k_m}$ consists of all non-negative solutions to $\sum_{m=1}^jmk_m = j$, $r\triangleq\sum_{m=1}^j k_m$. The set $\Brace{k_l}$ and $r$ in \eqref{eq: 17} are defined analogously.
\end{theorem}
\begin{remark}
 In Theorem \ref{theorem: 3}, the Cram\'er's condition holds if the random variable $W$ has a proper density function. The polynomial $He_j(x)$ is known as the Hermite polynomial of degree $j$. $\kappa_m$ denotes the order-$m$ cumulant of random variable $W/\sigma$. \eqref{eq: 17} indicates that $\kappa_m$ is a homogeneous polynomial in moments of degree $m$. In \cite{Blinnikov1998}, the authors presented a proof of \eqref{eq: 16} and \eqref{eq: 17} and provided an efficient algorithm to compute the set $\Brace{k_m}$ in \eqref{eq: 15}.
\end{remark}

As an application of Theorem \ref{theorem: 3}, let $W = 1-\log\big(1 + e^{-2P-2Z\sqrt{P}}\big) - C$, where $Z\sim\N(0,1)$. Clearly, $W$ has a proper density function and $\E[|W|^{s+2}] < \infty$ holds for any $s\in\Z_+$. Hence, for moderate and large values of $n$, the differentiable function $F_{\gamma}(n)$ we use to approximate $\Prob[\iota(X^n; Y^n) \ge \gamma]$ is given by the order-$s$ Edgeworth expansion, i.e.,
\begin{align}
&F_{\gamma}(n)\notag\\
&=Q\Paren{\frac{\gamma-nC}{\sqrt{nV}}} - \phi\Paren{\frac{\gamma-nC}{\sqrt{nV}}}\sum_{j=1}^{s}n^{-\frac{j}{2}}p_j\Paren{\frac{\gamma-nC}{\sqrt{nV}}}. \label{eq: Edgeworth expansion}
\end{align}

A caveat of using the order-$s$ Edgeworth expansion is that for small values of $n$, the order-$s$ Edgeworth expansion oscillates around $0$ due to truncation of an infinite series, making it no longer a suitable approximation function to the tail probability. To remedy the situation, we resort to the Petrov expansion \cite{Petrov1975} for small $n$.
\begin{theorem}[Theorem 1, \cite{Petrov1975}]
    Let $W_1, W_2, \dots, W_n$ be a sequence of i.i.d. random variables with zero mean and a finite variance $\sigma^2$. Define $G_n(x)\triangleq \Prob\Br{\sum_{i=1}^nW_i\le x\sigma\sqrt{n} }$. If $x\ge0$, $x = o(\sqrt{n})$, and the moment generating function $\E[e^{tW}]<\infty$ for $|t|<H$ for some $H>0$, then
    \begin{align}
    &G_n(x)=1-Q(x)\exp\Brace{\frac{x^3}{\sqrt{n}}\Lambda\Paren{\frac{x}{\sqrt{n}}} }\Br{1 + O\Paren{\frac{x+1}{\sqrt{n} } } },
    \end{align}
    \begin{align}
    &G_n(-x)= Q(x) \exp\Brace{\frac{-x^3}{\sqrt{n}}\Lambda\Paren{\frac{-x}{\sqrt{n}}} }\Br{1 + O\Paren{\frac{x+1}{\sqrt{n} } } },
    \end{align}
    where $\Lambda(t) = \sum_{k=0}^\infty a_kt^k$ is called the Cram\'er series\footnote{Details on Cram\'er series can be found in the proof of \cite[Theorem 2]{Petrov1975}.} .
\end{theorem}
In \cite{Petrov1975}, Petrov provided the order-$2$ Cram\'er series $\Lambda^{[2]}(t)$,
\begin{align}
    &\Lambda^{[2]}(t) \notag\\
    &= \frac{\kappa_3}{6\kappa_2^{3/2}} + \frac{\kappa_4\kappa_2 - 3\kappa_3^2}{24\kappa_2^3}t + \frac{\kappa_5\kappa_2^2 - 10\kappa_4\kappa_3\kappa_2 + 15\kappa_3^3 }{120\kappa_2^{9/2} }t^2. \label{eq: Cramer series}
\end{align}
For small $n$ satisfying $n < \gamma/C$, the function $F_{\gamma}(n)$ we use to approximate $\Prob[\iota(X^{n}; Y^{n}) \ge \gamma]$ is given by the order-$3$ Petrov expansion, where the order of $3$ is determined by $\kappa_5$ in \eqref{eq: Cramer series},
\begin{align}
    F_{\gamma}(n) = Q\Paren{\frac{\gamma-nC}{\sqrt{nV} } }\exp\Brace{\frac{(\gamma-nC)^3}{n^2V^{3/2} }\Lambda^{[2]}\Paren{\frac{\gamma-nC}{n\sqrt{V}} } }.
\end{align}

\begin{figure}[t]
\centering
\includegraphics[width=0.45\textwidth]{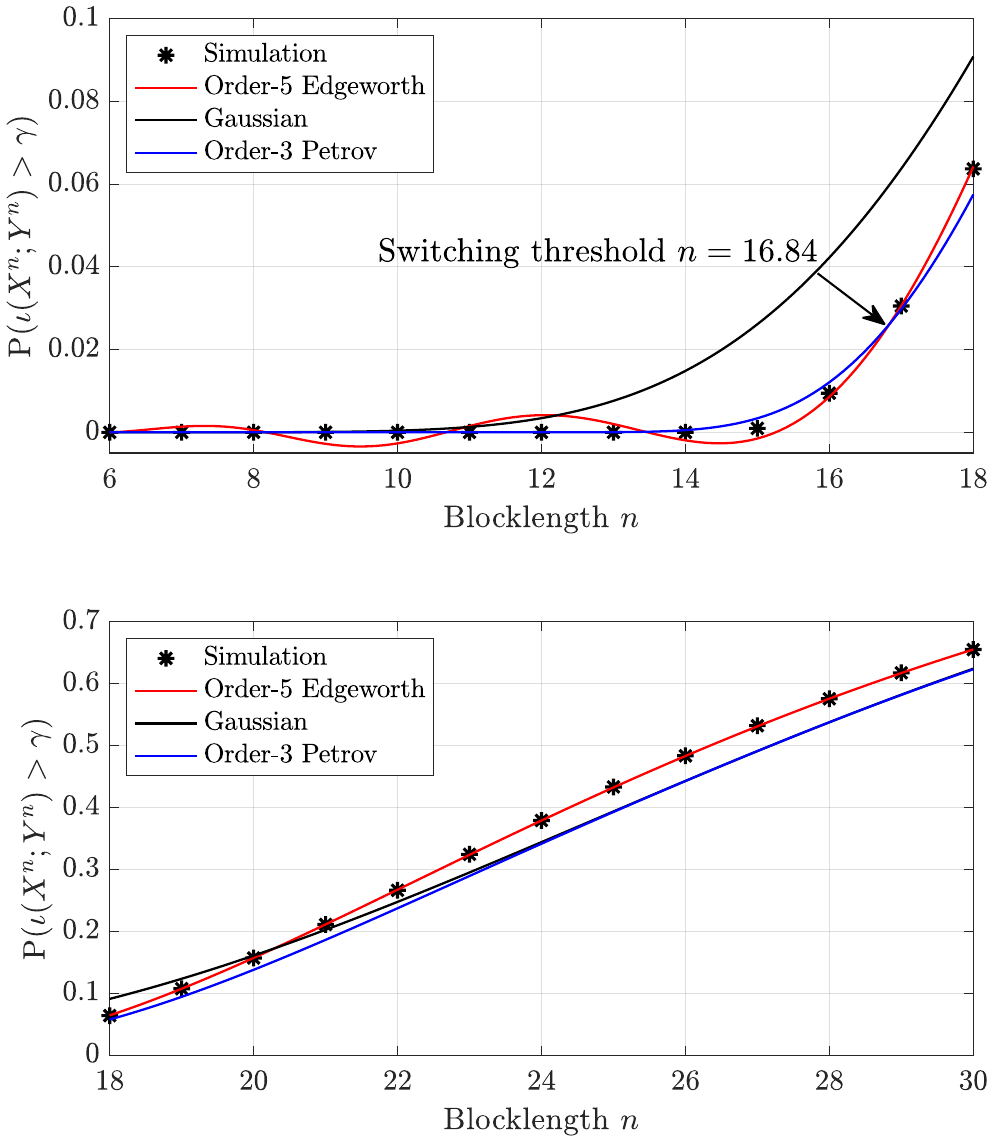}
\caption{Comparison of various approximation models for $\Prob[\iota(X^n; Y^n)\ge \gamma]$ with a fixed $\gamma > 0$. In this example, $k = 6$, $\epsilon = 10^{-2}$, $\gamma = \log\frac{2^k-1}{\epsilon/2} = 13.62$ for BI-AWGN channel at $0.2$ dB. }
\label{fig: approx models}
\end{figure}

In our implementation, we found that the order-$5$ Edgeworth expansion meets our desired approximation accuracy at large $n$. The switch from the order-$5$ Edgeworth expansion to the order-$3$ Petrov expansion occurs at the largest value for which two expansions are equal with a common value less than $1/2$. Fig. \ref{fig: approx models} shows the comparison of different approximation models for $\Prob[\iota(X^n; Y^n)\ge \gamma]$ with $\gamma = 13.62$ for BI-AWGN channel at $0.2$ dB. The Gaussian model in Fig. \ref{fig: approx models} is given by $Q\Paren{\frac{\gamma - nC}{\sqrt{nV}}}$. Fig. \ref{fig: approx models} shows that the Gaussian model fails to capture the true tail probability at small $n$. The order-5 Edgeworth expansion oscillates around $0$ when $n<16$ and is extremely accurate when $n\ge 16$. In contrast, the order-$3$ Petrov expansion is loose yet close to the Gaussian model when $n\ge 24$ and becomes tight when $n \le 14$. Therefore, the combination of the order-$2$ Petrov expansion and the order-$5$ Edgeworth expansion at switching threshold $n = 16.84$ provides a remarkably precise estimate of the tail probability $\Prob[\iota(X^n; Y^n)\ge \gamma]$.

\section{VLSF Codes With $m$ Optimal Decoding Times}\label{sec: SDO for VLSF codes}

In this section, we develop numerical tools to evaluate the achievable rate of a VLSF code with $m$ optimal decoding times. We mainly consider the error regime where Polyanskiy's scheme of stopping at zero does not improve the achievability bound \cite{Polyanskiy2011}.

\subsection{An Integer Program and a Greedy Algorithm}

In \cite{Yavas2021}, Yavas \emph{et al.} proved an achievability bound for an $(l, n_1^m, M, \epsilon)$ VLSF code for the AWGN channel. With a slight modification, this result is directly applicable to the  BI-AWGN channel.

\begin{theorem}[Theorem 3, \cite{Yavas2021}]\label{theorem: 1}
    Fix a constant $\gamma > 0$ and decoding times $n_1 < \cdots < n_m$. For any positive numbers $l$ and $\epsilon\in(0, 1)$, there exists an $(l, n_1^m, M, \epsilon)$ VLSF code for the BI-AWGN channel \eqref{eq: BI-AWGN} with
    \begin{align}
        &\E[\tau] \le n_1 + \sum_{i=1}^{m-1}(n_{i+1} - n_i)\Prob\Bigg[\bigcap_{j=1}^i\{\iota(X^{n_j}; Y^{n_j}) < \gamma \} \Bigg], \label{eq: 10}\\
      &P_e \le \Prob[\iota(X^{n_m}; Y^{n_m}) < \gamma] + (M-1)2^{-\gamma},  \label{eq: 11}
    \end{align}
    where $P_{X^{n_m}}$ is the product of distribution of $m$ subvectors of length $n_j - n_{j-1}$, $j\in[m]$, with the convention $n_0 = 0$. Namely,
    \begin{align}
    P_{X^{n_m}}(x^{n_m}) = \prod_{j=1}^mP_{X_{n_{j-1}+1}^{n_j}}(x_{n_{j-1}+1}^{n_j}).  \label{eq: codebook distribution}
    \end{align}
\end{theorem}

\begin{remark}
In \cite{Yavas2021} (and its full version \cite{Yavas_arXiv2021}), Yavas et al. obtained Theorem \ref{theorem: 1} by constructing a random VLSF code according to distribution \eqref{eq: codebook distribution}  and applying an information density threshold decoder that favors the largest message index whose cumulative information density exceeds $\gamma$ for the first time among any other message indices at decoding times $\{n_1, n_2, \dots,n_m\}$. 
 
 In \eqref{eq: 11}, the first term upper bounds the probability that the true message never crosses $\gamma$ and the second term upper bounds the probability that any other message crosses $\gamma$ sooner than the true message.
\end{remark}

Interested readers can refer to the full version \cite{Yavas_arXiv2021} of \cite{Yavas2021} for the proof of Theorem \ref{theorem: 1}. For our purposes, Theorem \ref{theorem: 1} motivates the following integer program. Define
\begin{align}
    &N(\gamma, n_1^m) \triangleq n_1 + \sum_{i=1}^{m-1}(n_{i+1} \,{-}\,n_i)\Prob[\iota(X^{n_i}; Y^{n_i}) < \gamma],\\
    &\F_m(\gamma, M, \epsilon) \triangleq\{n_1^m: n_{i+1} - n_i\ge1, i\in[m-1], \text{ and }\notag\\
    &\phantom{\F_m(\gamma, M)} \Prob[\iota(X^{n_m}; Y^{n_m}) < \gamma] + (M-1)2^{-\gamma} \le \epsilon \}. \label{eq: 18}
\end{align}
For a given $m\in\Z_+, M\in\Z_+$, $\epsilon\in(0, 1)$, and $\gamma\ge \log\frac{M-1}{\epsilon}$,
\begin{align}
\begin{split}
    \min&\quad N_{\gamma}(n_1^m) \\
  \st&\quad n_1^m\in\F_m(\gamma, M, \epsilon) \\
        &\quad n_1^m\in\Z_+^m.
\end{split}
\label{eq: integer-valued program}
\end{align}

In the integer program \eqref{eq: integer-valued program}, we consider the minimum gap and average error probability constraints  as in \eqref{eq: 18}, and the constraint that all decoding times must be integers.  

Let $\tilde{N}(\gamma)$ denote the locally minimum upper bound $N(\gamma, n_1^m)$ on $\E[\tau]$ for a given $\gamma$ in program \eqref{eq: integer-valued program}. Then, $\min_{\gamma}\tilde{N}(\gamma)$ yields the globally minimum upper bound $N(\gamma, n_1^m)$. In this paper, we solve the globally minimum upper bound $N(\gamma, n_1^m)$ using this two-step minimization approach.

In general, an integer program  is NP-complete. For the specific integer program \eqref{eq: integer-valued program}, additional challenge is caused by the fact that there is no closed-form expression for $\Prob[\iota(X^{n_k}; Y^{n_k}) < \gamma]$ and $n_1,n_2,\dots,n_m$ are required to be monotonically increasing integers. While a complete solution to the integer program \eqref{eq: integer-valued program} remains open, we establish the following results. 

\begin{lemma}\label{lemma: 1}
   Fix $m\in\Z_+, M\in\Z_+, \epsilon\in(0,1)$ and $\gamma\ge\frac{M-1}{\epsilon}$, and let $n_0\triangleq0$. Let $n_1^m\in\F_m(\gamma, M, \epsilon)$ be $m$ integer-valued decoding times that achieve $N(\gamma, n_1^m)$. Suppose there exists an integer $\tilde{n}$ such that $n_{k-1}< \tilde{n} < n_k$ for some $k\in[m]$. Then,
   \begin{align}
    N(\gamma, n_1^{k-1},\tilde{n},n_{k}^m) < N(\gamma,n_1^m). \label{eq: ineq_N}
   \end{align}
\end{lemma}

\begin{IEEEproof}
For brevity, let $S_n \triangleq \iota(X^{n}; Y^{n})$. Let $n_1^m\in\F_m(\gamma, M, \epsilon)$ be the $m$ integer-valued decoding times that achieve $N(\gamma, n_1^m)$. If $k=1$, i.e., $\tilde{n}<n_1$, then \eqref{eq: ineq_N} holds trivially. Assume that $2\le k\le m$ and there exists an integer $\tilde{n}$ such that $n_{k-1} < \tilde{n} < n_k$. It is straightforward to show that
    \begin{align}
    &N(\gamma, n_1^m) - N(\gamma, n_1^{k-1},\tilde{n},n_{k}^m) \notag\\
    &= (n_k - n_{k-1})\Prob[S_{n_{k-1}}<\gamma] - (\tilde{n}-n_{k-1})\Prob[S_{n_{k-1}}<\gamma] \notag\\
    &\phantom{--} - (n_k-\tilde{n})\Prob[S_{\tilde{n}}<\gamma] \ge 0, \notag
    \end{align}
    where the last step follows from $\Prob[S_{\tilde{n}}<\gamma] < \Prob[S_{n_{k-1}}<\gamma]$.
\end{IEEEproof}
Define
\begin{align}
    &N_{m}^*(\gamma,M,\epsilon) \triangleq \min_{n_1^m\in\F_m(\gamma, M, \epsilon)} N(\gamma, n_1^m).
\end{align}
\begin{theorem}\label{theorem: 2}
    Fix $M\in\Z_+$, $\epsilon\in(0, 1)$ and $\gamma\ge \log\frac{M-1}{\epsilon}$. Let $n^* = \min\{n\in\Z_{+}: \Prob[\iota(X^{n}; Y^{n}) < \gamma] \le \epsilon - (M-1)2^{-\gamma}\}$. For $m< n^*$, it holds that
    \begin{align}
        N_{m+1}^*(\gamma,M,\epsilon) < N_{m}^*(\gamma,M,\epsilon).
    \end{align}
\end{theorem}
\begin{IEEEproof}
    Let $n_0\triangleq0$. Let $n_1^m\in\F_m(\gamma, M, \epsilon)$ be the sequence of decoding times that achieves $N_{m}^*(\gamma,M,\epsilon)$. It is straightforward to show that the optimal $n_m = n^*$. If $m < n^*$, this implies that there exists some $k\in[m]$ such that $n_k - n_{k-1}\ge2$. For this $k$, choose an integer $\tilde{n}$ such that $n_{k-1} < \tilde{n} < n_k$. Clearly, $(n_1^{k-1}, \tilde{n}, n_k^m)\in\F_{m+1}(\gamma, M, \epsilon)$. Therefore, 
    \begin{align}
        N_{m+1}^*(\gamma,M,\epsilon) \le N(\gamma, n_1^{k-1}, \tilde{n}, n_k^m) &<  N_{m}^*(\gamma, M, \epsilon),\label{eq: ineq_19}
    \end{align}
    where the last inequality in \eqref{eq: ineq_19} follows from Lemma \ref{lemma: 1}.
\end{IEEEproof}
Theorem \ref{theorem: 2} motivates the following greedy algorithm for a fixed $\gamma$: Start from $m = n^*$ where $n^* \triangleq \min\{n\in\Z_+: \Prob[\iota(X^n; Y^n)<\gamma]\le \epsilon-(M-1)2^{-\gamma}\}$. Suppose that $n_1^m$ is the solution for $m$. Then, the solution $\tilde{n}_1^{m-1}$ for $m-1$ is identified by removing the decoding time $n_i$ in $n_1^{m-1}$ that minimizes $N_{\gamma}(n_1^{i-1}, n_{i+1}^{m})$. Note that the decoding time $n_m$ is always retained to ensure that the target error probability is met via \eqref{eq: 11}.



\subsection{The Relaxed Program and the Gap-Constrained SDO}

To facilitate a program that is computationally tractable, we consider the relaxed program that allows $n_1^m\in\R_+^m$: For a given $m\in\Z_+, M\in\Z_+$, $\epsilon \in (0, 1)$ and $\gamma \ge \log\frac{M-1}{\epsilon}$,
\begin{align}
\begin{split}
    \min&\quad N(\gamma, n_1^m) \\
  \st&\quad n_1^m\in\F_m(\gamma, M, \epsilon),
\end{split}
\label{eq: real-valued program}
\end{align}
where the tail probability $\Prob[\iota(X^{n}; Y^{n}) \ge \gamma]$ is approximated by a monotonically increasing and differentiable function $F_{\gamma}(n)$ with $F_{\gamma}(0) = 0$ and $F_{\gamma}(\infty) = 1$, for instance, the piecewise function\footnote{The first derivative of $F_{\gamma}(n)$ at the switching threshold does not exist. Nonetheless, one can assign the right (or left) derivative as the derivative for the switching threshold so that the solution is not affected significantly.} introduced in Section \ref{sec: approximation}. Let 
\begin{align}
f_{\gamma}(n) \triangleq \frac{dF_{\gamma}(n)}{dn}. 
\end{align}

For the relaxed program \eqref{eq: real-valued program} with a fixed $\gamma$, the optimal, real-valued decoding times $n_1^*, n_2^*, \dots, n_m^*$ are given by the following theorem.
\begin{theorem}
   For a given $m\in\Z_+, M\in\Z_+$, $\epsilon \in (0, 1)$ and $\gamma \ge \log\frac{M-1}{\epsilon}$, the optimal real-valued decoding times $n_1^*, n_2^*, \dots, n_m^*$ in program \eqref{eq: real-valued program} satisfy
   \begin{align}
        &n_m^* = F_{\gamma}^{-1}\Paren{1 - \epsilon + (M-1)2^{-\gamma} },  \label{eq: 20} \\
    & n_{k+1}^* = n_k^* + \max\Brace{1,\frac{F_{\gamma}(n_k^*) - F_{\gamma}(n_{k-1}^*) - \lambda_{k-1} }{f_{\gamma}(n_k^*)}}, \label{eq: 21} \\
    & \lambda_k = \max\{\lambda_{k-1}+f_{\gamma}(n_k^*) - F_{\gamma}(n_k^*) + F_{\gamma}(n_{k-1}^*), 0\}, \label{eq: 22}
\end{align}
   where $k\in[m-1]$, $\lambda_0 \triangleq 0$ and $n_0^*\triangleq 0$.
\end{theorem}

\begin{IEEEproof}
For brevity, define $\bm{n} \triangleq (n_1, n_2, \dots, n_m)$. By introducing the Lagrangian multipliers $\nu$, $\lambda_1^{m-1}$, the Lagrangian of program \eqref{eq: real-valued program} is given by
\begin{align}
    &\LL(\bm{n}, \nu, \lambda_1^{m-1}) = n_1 + \nu(1 - F_{\gamma}(n_m) + (M-1)2^{-\gamma} - \epsilon)  \notag\\
    &\phantom{=} + \sum_{i=1}^{m-1}(n_{i+1} - n_i)(1 - F_{\gamma}(n_i) )+ \sum_{i=1}^{m-1}\lambda_i(n_i - n_{i+1} + 1). \notag
\end{align}
By the Karush-Kuhn-Tucker (KKT) conditions, the optimal decoding times $\bm{n}^* = (n_1^*, n_2^*,\dots, n_m^*)$ must satisfy
\begin{align}
    &\frac{\partial \LL}{\partial n_k}\Big|_{\bm{n}=\bm{n}^*} = F_{\gamma}(n_k^*) - F_{\gamma}(n_{k-1}^*) - (n_{k+1}^* - n_k^*)f_{\gamma}(n_k^*) \notag\\
    &\phantom{\frac{\partial \LL}{\partial n_k}\Big|_{\bm{n}=\bm{n}^*} =}+ \lambda_k - \lambda_{k-1} = 0,\quad k\in[m-1], \label{eq: 23}\\
    &\frac{\partial \LL}{\partial n_m}\Big|_{\bm{n}=\bm{n}^*} = 1 - F_{\gamma}(n_{m-1}^*) - \nu f_{\gamma}(n_m^*) = 0, \label{eq: 24}\\
    &\nu(1 - F_{\gamma}(n_m^*) +(M-1)2^{-\gamma}-\epsilon) = 0, \label{eq: 25} \\
  &\lambda_k(n_k^* - n_{k+1}^* + 1) = 0,\quad k\in[m-1]. \label{eq: 26}
\end{align}
Since $F_{\gamma}(n)\in(0,1)$ and $f_{\gamma}(n) > 0$ for $n>0$, \eqref{eq: 24} indicates that $\nu > 0$. Hence, we obtain $n_m^* = F_{\gamma}^{-1}\Paren{1 - \epsilon + (M-1)2^{-\gamma} }$ from \eqref{eq: 25}. 

Next, we analyze \eqref{eq: 26}. There are two cases. If $\lambda_k > 0$, then $n_{k+1}^* = n_k^* + 1$. By \eqref{eq: 23}, we obtain
\begin{align}
    \lambda_k = \lambda_{k-1}+f_{\gamma}(n_k^*) - F_{\gamma}(n_k^*) + F_{\gamma}(n_{k-1}^*).
\end{align}
If $n_{k+1}^* > n_k^* + 1$, then $\lambda_k = 0$. By \eqref{eq: 23}, we obtain
\begin{align}
    n_{k+1}^* = n_k^* + \frac{F_{\gamma}(n_k^*) - F_{\gamma}(n_{k-1}^*) - \lambda_{k-1} }{f_{\gamma}(n_k^*)}.
\end{align}
Rewriting the above two cases in a compact form yields \eqref{eq: 21} and \eqref{eq: 22}.
\end{IEEEproof}

The procedures \eqref{eq: 21} and \eqref{eq: 22} are called the \emph{gap-constrained SDO} for the relaxed program \eqref{eq: real-valued program}. In contrast, the SDO studied in \cite{Vakilinia2016,Wang2017,Wesel2018,Wong2017,Heidarzadeh2018,Heidarzadeh2019} is derived from the relaxed program \eqref{eq: real-valued program} without the gap constraint\footnote{The error probability constraint is also different, yet it does not affect the SDO procedure.} and admits a simple recursion
\begin{align}
    n_{k+1}^* = n_k^* + \frac{F_{\gamma}(n_k^*) - F_{\gamma}(n_{k-1}^*) }{f_{\gamma}(n_k^*)},\quad k\in[m-1], \label{eq: SDO without gap constraint}
\end{align}
where $n_0^* \triangleq 0$. We will show that for small values of $m$, the gap-constrained SDO behaves indistinguishably as the SDO without the gap constraint in \eqref{eq: SDO without gap constraint}. However, as $m$ becomes large, the decoding times provided by these two algorithms differ noticeably.

In practice, after solving $n_m^*$ via \eqref{eq: 20}, one would apply a bisection search between $0.5$ and $\lceil n_m^*\rceil - m + 0.5$ for $n_1$ and the SDO to identify $n_1^*$. This guarantees that the nearest integer to $n_1^*$ is at least $1$.

When evaluating at small $n$, both $F_{\gamma}(n)$ and $f_{\gamma}(n)$ will become infinitesimally small. In this case, a direct numerical computation using \eqref{eq: 21} and \eqref{eq: 22} may cause the precision issue. Fortunately, the SDO described by \eqref{eq: 21} and \eqref{eq: 22} also admits a ratio form. Define $\lambda_k^{(r)} \triangleq \lambda_k / f_{\gamma}(n_k^*) $. Thus, \eqref{eq: 21} and \eqref{eq: 22} are equivalent to
\begin{align}
    & n_{k+1}^*\,{=}\,n_k^*+\max\Big\{1, \frac{F_{\gamma}(n_k^*)}{f_{\gamma}(n_k^*)} {-} \frac{F_{\gamma}(n_{k-1}^*)}{f_{\gamma}(n_k^*)} {-} \lambda_{k-1}^{(r)}\frac{f_{\gamma}(n_{k-1}^*)}{f_{\gamma}(n_k^*)} \Big\}, \notag \\
    & \lambda_k^{(r)}\,{=}\,\max\Big\{\lambda_{k-1}^{(r)}\frac{f_{\gamma}(n_{k-1}^*)}{f_{\gamma}(n_k)^*}\,{+}\,1\,{-}\,\frac{F_{\gamma}^*(n_k)}{f_{\gamma}^*(n_k)}\,{+}\,\frac{F_{\gamma}^*(n_{k-1})}{f_{\gamma}^*(n_k)}, 0\Big\}.  \notag
\end{align}
The purpose of using $F_{\gamma}(\tilde{n})/f_{\gamma}(n)$, $f_{\gamma}(\tilde{n})/f_{\gamma}(n)$, and $\lambda_k^{(r)}$ is that they have a closed-form expression that cancels out the common infinitesimal factor in both the numerator and denominator. In our implementation, we applied the ratio form of SDO.

\subsection{Error Regime Where Stopping at Zero Does Not Help}

In \cite{Polyanskiy2011}, Polyanskiy \emph{et al.} demonstrated that the VLSF code with infinitely many stopping times can achieve $\frac{C}{1-\epsilon}$. This is accomplished by the following scheme: With probability $p = \frac{\epsilon - \epsilon'}{1 - \epsilon'}$, the code immediately stops at $\tau = 0$ without any channel use, and with probability $1 - p$, employs an $(l', M, \epsilon')$ VLSF code satisfying $\log M = Cl' + \log \epsilon' - a_0$, where $a_0 \triangleq \sup_{x,y}\iota(x;y)$. The overall code has an error probability
\begin{align}
    1\cdot p + \epsilon'(1 - p) = \epsilon,
\end{align}
and average blocklength
\begin{align}
    0\cdot p +  l'(1 - p) = l'(1-p).
\end{align}

In this section, we identify the error regime where Polyanskiy's scheme of stopping at $\tau = 0$ does not improve the achievability bound. 
\begin{theorem}\label{theorem: stopping does not help}
    For a given $a_0\in\R_{+}$, $M\in\Z_{+}$, define
    \begin{align}
        \epsilon^* \triangleq \argmin_{x\in(0, 1)}\frac{\log M + a_0 - \log x}{1 - x}.
    \end{align}
    If  $\epsilon\in(0, \epsilon^*]$, stopping at $\tau = 0$ does not improve the achievability bound for VLSF codes.
\end{theorem}
\begin{IEEEproof}
    By Polyanskiy's scheme, solving the error regime where stopping at zero does not improve the achievability bound is equivalent to identifying the error regime in which $\epsilon' = \epsilon$ is the minimizer to the following program: For a given $C, a_0\in\R_{+}$, $M\in\Z_{+}$, and  $\epsilon\in(0, 1)$,
\begin{align}
\begin{split}
    \min_{\epsilon'}&\quad l'(1 - p) \\
    \st&\quad \log M = Cl' + \log \epsilon' - a_0 \\
        &\quad p = \frac{\epsilon - \epsilon'}{1 - \epsilon'} \\
        &\quad \epsilon'\in(0, \epsilon].
\end{split}
\label{eq: Polyanskiy program}
\end{align}
The program \eqref{eq: Polyanskiy program} is equivalent to the following program
\begin{align}
\begin{split}
    \min_{\epsilon'}&\quad \Paren{\frac{1 - \epsilon}{C} }f(\epsilon') \\
    \st&\quad \epsilon'\in(0, \epsilon],
\end{split}
\label{eq: simple Polyanskiy program}
\end{align}
where 
\begin{align}
    f(x)\triangleq \frac{\log M + a_0 - \log x}{1 - x}. 
\end{align}
Since $f(x)$ is convex in $(0, 1)$, there exists a unique minimizer $\epsilon^*\in(0, 1)$. Therefore, if $\epsilon \le \epsilon^*$, then $\epsilon' = \epsilon$ minimizes the objective function in \eqref{eq: simple Polyanskiy program}, giving $p = 0$. Namely, stopping at zero does not improve the achievability bound.
\end{IEEEproof}
We remark that there is no closed-form solution to $\epsilon^*$ in Theorem \ref{theorem: stopping does not help}. Nonetheless, one can numerically solve $\epsilon^*$ for a given $M$ and $a_0$.

\begin{figure}[t]
\centering
\includegraphics[width=0.45\textwidth]{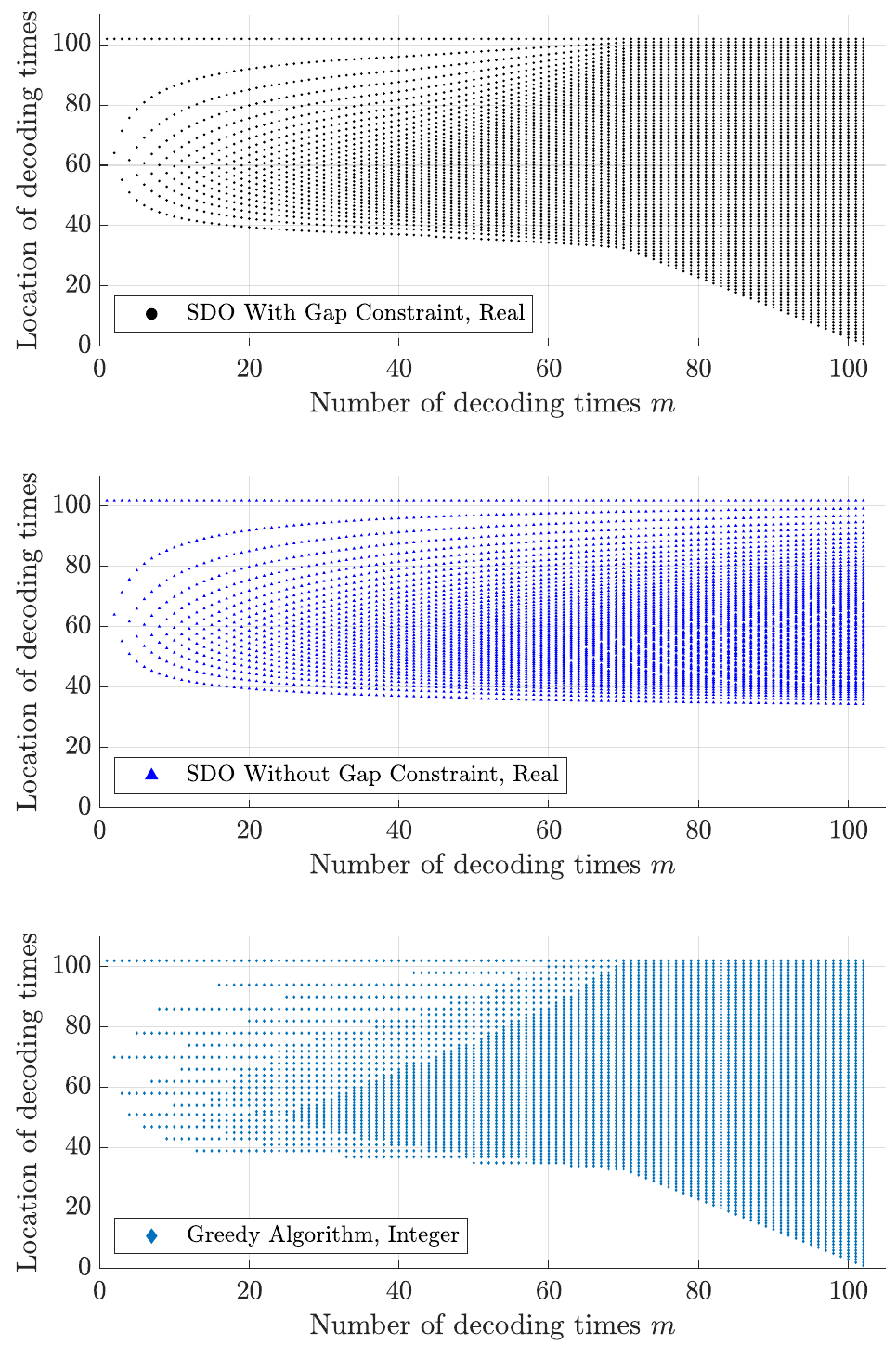}
\caption{Comparison of the real-valued decoding times by the SDO with gap constraint, the real-valued decoding times by the SDO without gap constraint, and the integer-valued decoding times by greedy algorithm for $k = 20$, $\epsilon = 10^{-2}$, $\delta = 1/2$, $\gamma = \log\frac{2^k-1}{\delta\epsilon}$ and BI-AWGN channel at $0.2$ dB. $m$ ranges from $1$ to $\lceil n_m^*\rceil = 102$, where $n_m^* = 101.91$ is given by \eqref{eq: 20}. }
\label{fig: dec time evolution}
\end{figure}

\begin{figure}[t]
\centering
\includegraphics[width=0.45\textwidth]{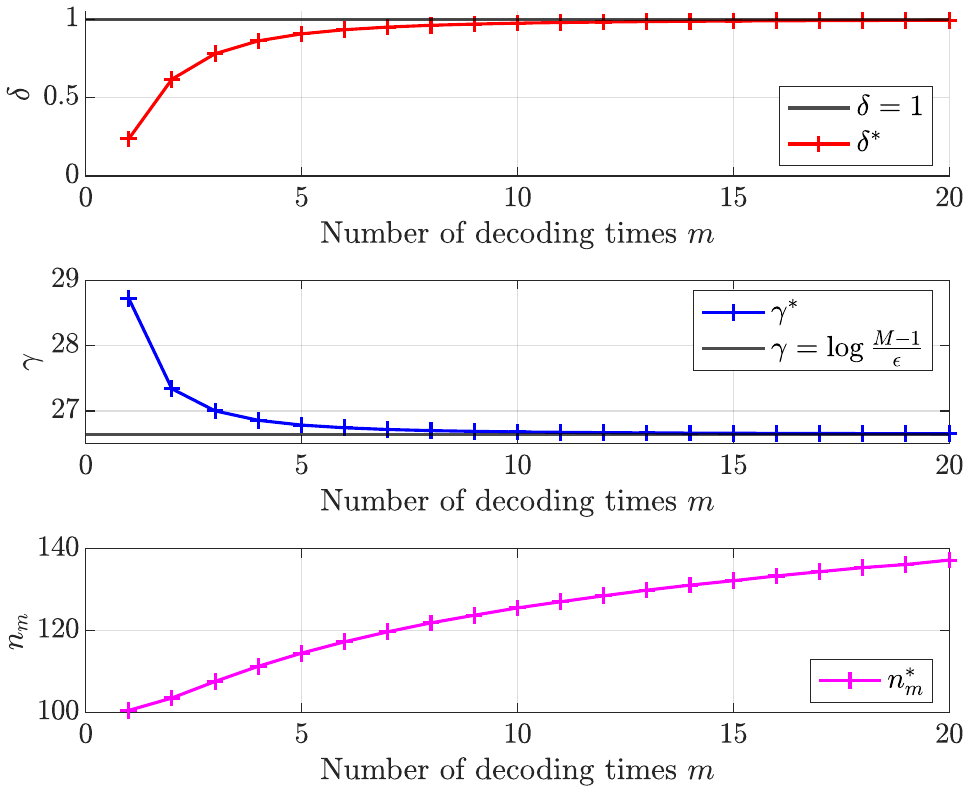}
\caption{Globally optimal $\delta^*$, $\gamma^*$, and $n_m^*$ as a function of the number of decoding times $m$. In the VLSF achievability bound, $\delta = 1$ and $\gamma = \log\frac{M-1}{\epsilon}$. In this example, $\epsilon = 10^{-2}$, $k = 20$ for BI-AWGN channel at $0.2$ dB. }
\label{fig: opt_delta_gamma_vs_m}
\end{figure}

\subsection{Numerical Evaluation}
Let $M = 2^k$, $k\in\Z_{+}$. We consider the BI-AWGN channel at $0.2$ dB with a capacity of $0.5$ and the error regime in which stopping at zero does not improve the achievability bound. By Theorem \ref{theorem: stopping does not help}, if $k\le 100$, $\epsilon\le 1.33\cdot10^{-2}$ is the error regime where stopping at zero does not help. In the following example, we consider $\epsilon = 10^{-2}$.  

We consider the relaxed program and apply the two-step minimization with the gap-constrained SDO introduced in Section \ref{sec: SDO for VLSF codes} to obtain the globally minimum upper bound $N^*(\gamma, n_1^m)$. Thus, $k/N^*(\gamma, n_1^m)$ gives the achievability bound.  In \cite{Polyanskiy2011}, Polyanskiy \emph{et al.} showed that the average blocklength $\E[\tau]$ of a VLSF code with $m = \infty$ and no stopping at $\tau=0$ is upper bounded by
\begin{align}
    \E[\tau]\le \frac{\log\frac{M-1}{\epsilon} + a_0}{C}, \label{eq: VLSF bound}
\end{align}
where $a_0 \triangleq \sup_{x,y}\iota(x;y)$. This bound yields the \emph{VLSF achievability bound} on rate. For BI-AWGN channel, $a_0 = 1$.

For a fixed $\gamma$ at $k = 20$ and $\epsilon = 10^{-2}$, Fig. \ref{fig: dec time evolution} shows how the decoding times evolve with $m$ for the three algorithms: SDO with/without the gap constraint, and the greedy algorithm.  For $m\le 20$, SDO with a gap constraint behaves indistinguishably as the SDO without a gap constraint since the SDO solution naturally has gaps larger than one. The greedy algorithm is forced to choose from the remaining decoding times, leading to a possibly suboptimal solution. For large $m$, SDO without the gap constraint avoids early decoding times and instead adds later decoding times so densely that their separation is less than one. In contrast, SDO with the gap constraint is forced to add early decoding times when all existing gaps become one. 

The greedy algorithm lacks the optimality guarantee of the gap-constrained SDO and is computationally more intensive.  Despite their distinct design perspectives, the greedy algorithm and the gap-constrained SDO arrive at essentially the same solution for large $m$. For $k = 20$, $\epsilon = 10^{-2}$, and $\gamma = \log\frac{(2^k-1)}{\epsilon/2}$, Fig. \ref{fig: dec time evolution} shows that $n_1$ is never less than $37$ when $m\le 32$ and grows as the number of decoding times decreases.

We remark that Fig. \ref{fig: dec time evolution} assumes a constant $\gamma$ over all number of decoding times. However, in the two-step minimization with the gap-constrained SDO, the globally optimal $\gamma^*$ is a function of $m$. Therefore, the globally optimal $n_m^*$ may not stay as constant as shown in Fig. \ref{fig: dec time evolution}.

Let $\delta\in[0, 1]$ and assume that the first and second terms in the right-hand side of \eqref{eq: 11} are equal to $(1-\delta)\epsilon$ and $\delta\epsilon$, respectively. Then, both $\gamma^*$ and $n_m^*$ can be thought as a function of $\delta^*$, i.e.,
\begin{align}
&\gamma^*(\delta^*) = \log\frac{M-1}{\epsilon\delta^*}, \label{eq: 44} \\
& n_m^*(\delta^*) = F_{\gamma^*}^{-1}(1 - \epsilon + \epsilon\delta^*).\label{eq: 45} 
\end{align}
Thus, minimization over $\gamma$ is equivalent to minimization over $\delta$. For $k = 20$, $\epsilon = 10^{-2}$, and BI-AWGN channel at $0.2$ dB, Fig. \ref{fig: opt_delta_gamma_vs_m} shows how the globally optimal $\delta^*$ and the associated globally optimal $\gamma^*, n_m^*$ vary with $m$ during the two-step minimization. We see that when $m$ is small, $\delta^*$ is far from $1$, indicating a large value of $\gamma^*$ and a small value of $n_m^*$. As $m$ gets large, we observe that $\delta^*$ monotonically increases, which, by \eqref{eq: 44} and \eqref{eq: 45}, implies that $\gamma^*$ decreases and $n_m^*$ increases. In particular, as $m\to\infty$, $\delta^*\to 1$, and consequently,
\begin{align}
    &\lim_{\delta^*\to1}\gamma^*(\delta^*) = \log\frac{M-1}{\epsilon},\\
    &\lim_{\delta^*\to1}n_m^*(\delta^*) = \infty.
\end{align}
In Polyanskiy's setting \cite{Polyanskiy2011}, the first term in \eqref{eq: 11} is zero since $n_m = \infty$ and the optimal $\gamma$ can thus be computed as $\gamma = \log\frac{M-1}{\epsilon}$ from \eqref{eq: 11}, implying that $\delta = 1$. Fig. \ref{fig: opt_delta_gamma_vs_m} shows that as $m$ increases, $\delta^*$, $\gamma^*$, and $n_m^*$ rapidly  approach those in Polyanskiy's setting.


\begin{figure}[t]
\centering
\includegraphics[width=0.45\textwidth]{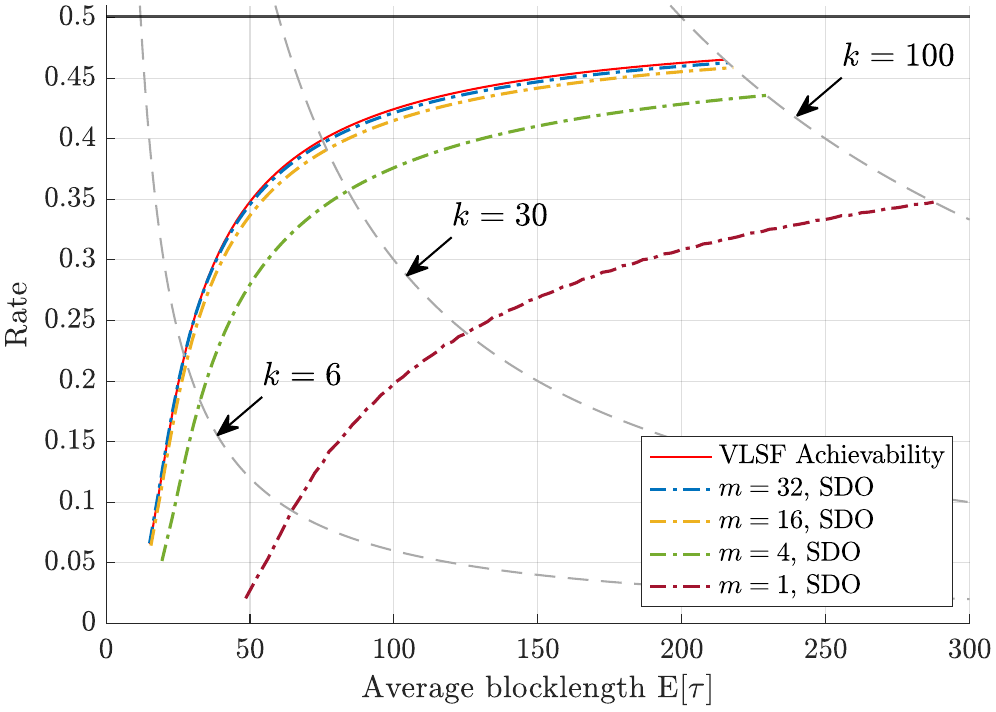}
\caption{Comparison of achievable rate estimation by the gap-constrained SDO and by the greedy algorithm for VLSF codes with $m$ optimal decoding times. The gray dashed line represents the $(\E[\tau], R)$ pairs such that $R\E[\tau] = k$. In this example, $\epsilon = 10^{-2}$ and the BI-AWGN channel is at $0.2$ dB.}
\label{fig: sim on BIAWGN}
\end{figure}

Fig. \ref{fig: sim on BIAWGN} shows the achievable rate of a VLSF code with $m$ optimal decoding times estimated by the two-step minimization with the gap-constrained SDO algorithm. We see that a finite $m$ suffices to achieve Polyanskiy's VLSF achievability bound derived from VLSF codes with infinitely many decoding times. For instance, for the BI-AWGN channel at $0.2$ dB and $\epsilon = 10^{-2}$, the achievable rate estimated by the SDO for VLSF codes with $k\le 6$ and $m = 32$ beats the VLSF bound. Additionally, for BI-AWGN channel at $0.2$ dB, with $16$ decoding times, the achievable rate by SDO is within $0.66\%$ of the VLSF achievability bound for $k \le 100$. With $32$ decoding times, it becomes hard to distinguish the achievable rate by SDO from the VLSF achievability bound for $k\le 30$.

\section{Conclusion}\label{sec: conclusion}
This paper provides a new SDO that includes the gap constraint.  Using this improved SDO, the paper demonstrates that Polyanskiy's VLSF achievability bound with infinitely many decoding times can be closely approached with a finite, and relatively small, number of decoding times. 


\balance
\bibliographystyle{IEEEtran}
\bibliography{IEEEabrv,references}

\end{document}